\documentclass[12pt]{iopart}
\usepackage{iopams}
\expandafter\let\csname equation*\endcsname\relax
\expandafter\let\csname endequation*\endcsname\relax
\usepackage{amsmath}
\usepackage{amssymb}
\usepackage{graphicx}
\usepackage{tensor}
\usepackage{setspace}
\usepackage{pzccal}
\usepackage{cite}
\usepackage{hyperref}
\def\mean#1{\left< #1 \right>} \def\FF{\mathpzc{F}}
\begin{document}
\title{Quasilocal energy and thermodynamic equilibrium conditions}
\author{Nezihe Uzun and David L. Wiltshire}
\address{Department of Physics and Astronomy, University of Canterbury,
Private Bag 4800, Christchurch 8140, New Zealand}
\date{Received 25 March 2015; Accepted 17 June 2015; Published 28 July 2015}
%
\begin{abstract}
Equilibrium thermodynamic laws are typically applied to horizons in general
relativity without stating the conditions that bring them into equilibrium. We
fill this gap by applying a new thermodynamic interpretation to a generalized
Raychaudhuri equation for a worldsheet orthogonal to a closed spacelike
2--surface, the ``screen'', which encompasses a system of arbitrary size in
nonequilibrium with its surroundings in general. In the case of spherical
symmetry this enables us to identify quasilocal thermodynamic potentials
directly related to standard quasilocal energy definitions. Quasilocal
thermodynamic equilibrium is defined by minimizing the mean extrinsic curvature
of the screen. Moreover, without any direct reference to surface gravity, we
find that the system comes into quasilocal thermodynamic equilibrium when the
screen is located at a generalized apparent horizon. Examples of the
Schwarzschild, Friedmann--Lema\^{\i}tre and Lema\^{\i}tre--Tolman geometries
are investigated and compared.
Conditions for the quasilocal thermodynamic and hydrodynamic equilibrium
states to coincide are also discussed, and a quasilocal virial relation is
suggested as a potential application of this approach.
\end{abstract}
\bigskip
\leftline{{\em Class.\ Quantum Grav.}\ {\bf32} (2015) 165011}
\maketitle

\section{Introduction}
A thermodynamic description of general relativity has been a long-sought goal \cite{1928PNAS...14..268T, 1933Sci....77..291T} which intensified with the advent of black hole mechanics \cite{PhysRevD.7.2333, PhysRevD.9.3292, Bardeen:1973gs}.
Most studies in the literature focus on {\em equilibrium} thermodynamics\footnote{See \cite{PhysRevLett.96.121301,PhysRevD.81.024016,Freidel:2013jfa,Freidel:2014qya} for some exceptions.} of {\em horizons}, without stating the conditions that bring them into equilibrium. In fact, in gravitational physics there are no well-defined conditions for defining equilibrium in terms of the behaviour of a {\em system} itself. In this paper we will take steps to remedy this by defining a
quasilocal thermodynamic equilibrium condition using a purely geometric
approach. We will focus on the extrinsic geometry of the closed spacelike
2--surface that appears both in various quasilocal energy definitions
\cite{1997GReGr..29..307K,Liu:2003bx,Epp:2000zr} and the generalized Raychaudhuri equation of
Capovilla and Guven \cite{Capovilla:1994yk}. This connection provides
a natural way of defining quasilocal thermodynamic potentials, and hence
equilibrium conditions.

In general, if one wants to investigate the energy exchange mechanisms of a
gravitational system from the thermodynamic viewpoint, the system should
have a finite spatial size. Energy definitions which refer to the spatial
asymptotic behaviour are not good candidates for general thermodynamic
equations. Thus quasilocal energy definitions, which refer to a Hamiltonian
on the 2-dimensional spacelike boundary
\cite{Brown:1992br,1997GReGr..29..307K}, are very important for general
relativistic thermodynamics. Here we will link such definitions to a
generalized notion of the work done in the deviation of worldsheet
congruences, to define quasilocal thermodynamic potentials. We also provide
a quasilocal first law by considering a {\em worldsheet total variation}, in
which a quasilocal temperature can be understood as a {\em worldsheet--constant}.

We will consider general spherically symmetric spacetimes, rather than just
black holes. Naturally, the thermodynamics of more general spacetimes is
also of interest. However, quasilocal definitions require closed surfaces on
which to integrate to define the system in question. Thus we believe that
the new approach taken here could also be applied in future to systems
whose bounding surfaces are defined by different symmetries.

Although early investigations dealt with equilibrium thermodynamics
of black hole event horizons, in the last few decades more general trapping, apparent and dynamical horizons of generic spacetimes have been introduced
\cite{Hayward:1993wb, Hayward:1997jp,Ashtekar:2003hk,Cai:2005ra,Faraoni:2011hf} .
Hayward's construction of equilibrium thermodynamics on trapping horizons \cite{Hayward:1993wb} highlights the significance of generalized apparent horizons.
While the original definition of apparent horizons applies to black holes
which require asymptotically flat spatial hypersurfaces \cite{Hawking:1973uf},
Hayward's generalized apparent horizon which was first constructed for black holes, has also been applied in more general cases \cite{Booth:2005qc,Faraoni:2011hf,Tian:2014sca}.
These include cosmological applications, where the generalized apparent horizon
is not necessarily spacelike but can be timelike or null depending on the
equation of state of the cosmic fluid \cite{Faraoni:2011hf}.

In this paper we will consider a spherically symmetric gravitational
system of arbitrary size which is not in equilibrium with its surroundings.
As one of our results we will show that when a particular equilibrium condition
is applied to such a system then the 2--surface enclosing the system is
located at the generalized apparent horizon of \cite{Hayward:1993wb}.
This result makes no direct reference to the surface gravity, which is
conventionally used to define the temperature of the horizon.

The paper is constructed as follows. In section~\ref{energydef} we review the quasilocal energy definitions in literature, highlighting those features important for our new construction, including in particular the determination of a well-defined internal energy.
Section~\ref{geoandequil} starts with a short summary of the geometric formalism developed by Capovilla and Guven \cite{Capovilla:1994yk} and continues with interpretation of their generalized Raychaudhuri equation as a thermodynamic relation.
Following this, a quasilocal thermodynamic equilibrium condition and the corresponding thermodynamic potentials are introduced.
In section ~\ref{examples} these results are applied to the Schwarzschild, Friedmann-Lema\^{\i}tre-Robertson-Walker and Lema\^{\i}tre-Tolman spacetimes.
In section~\ref{seclocalvsqlocal} we highlight the difference between local thermodynamics of matter fields on curved background and quasilocal gravitational thermodynamics, as a precursor to suggesting a potential application of our approach to a quasilocal virial relation in section~\ref{secvirial}.
We will use natural units in which $c, G, \hbar, k_B$ are taken to be 1 throughout the paper.
The metric signature will be $\left(-,+,+,+\right)$.
\section{Quasilocal energy definitions in literature}\label{energydef}
Here, we will discuss those quasilocal energy definitions which will be most relevant for our investigation of equilibrium thermodynamic conditions.
A detailed review of quasilocal energy definitions can be found in \cite{Szabados:2004vb}.
\subsection{Misner-Sharp-Hernandez (MSH) Energy}

In horizon thermodynamics, there is a broad consensus \cite{Hayward:1994bu,Cai:2006rs,Nielsen:2008kd,Wu:2009wp,Faraoni:2011hf} on the choice of the internal energy of a generic spherically symmetric spacetime.
It is usually taken as the Misner-Sharp-Hernandez energy \cite{Misner:1964je, Hernandez:1966zia} which we now define.
Consider a spherically symmetric spacetime metric with coordinates $\{ x^\alpha,\theta ,\phi \}$ where $\{x^\alpha\}=\{t,r\}$,
\begin{equation}
ds^2=\gamma _{\alpha \beta}dx^{\alpha}dx^{\beta}+R^2(x)\left(d\theta ^2 +\sin ^2\theta d\phi^2\right),
\end{equation}
$R$ being the areal radius.
In order to study time evolution, one can pick a preferred timelike vector, called the Kodama vector \cite{Kodama:1979vn}, which can be used to define surface gravity for dynamic spherically symmetric spacetimes \cite{Hayward:1994bu}.
Surface gravity, up to a constant, is in general related to the temperature defined on the horizon.
The Kodama vector is unique and it is parallel to the timelike Killing vector in static spacetimes.
Its components are given by,
\begin{eqnarray}
K^{\alpha}(x)=\epsilon ^{\alpha \beta}\partial _{\beta}R,\qquad
K^{\theta}=0,\qquad
K^{\phi}=0,
\end{eqnarray}
where $\epsilon _{\alpha \beta}$ is the Levi-Civita tensor in 2-dimensions.
Now consider a 3-dimensional spacelike hypersurface, $\Sigma$, with induced coordinates, $\{y^{\sigma}\}$, induced metric, $h_{\sigma \kappa}$, and unit normal, $n^{\nu}$, aligned with the Kodama vector.
The Kodama vector is associated with conserved charges including an energy
\begin{equation}
E_{MSH}= \int _{\Sigma}{d^3y\sqrt{h}T_{\mu \nu}K^{\mu}n^{\nu}},
\end{equation}
where $T_{\mu \nu}$ is the stress-energy tensor of matter in the 4-dimensional spacetime.
This defines the Misner-Sharp-Hernandez energy which can also be written
\begin{equation}
E_{MSH}=\frac{R}{2}\left(1-\gamma^{\alpha \beta}\partial _{\alpha}R\partial _{\beta}R\right).
\end{equation}
\subsection{Brown-York (BY) Energy}
Brown and York \cite{Brown:1992br} followed a Hamilton-Jacobi approach to define a quasilocal energy on a 2-dimensional spacelike boundary, $\mathbb{S} _t=\partial \Sigma _t$, of a compact spacelike 3-dimensional hypersurface, $\Sigma_t$.
These hypersurfaces are taken to foliate a compact spacetime domain, $D$, with topology $\Sigma \times [ t_1,t_2]$.
The fundamental object of the Brown-York definition is the 3-dimensional timelike boundary of the domain $D$ which will be denoted by $B$.
The boundary $B$ is foliated by $\mathbb{S}_t$ and it admits a fixed metric.
According to this approach, the stress-energy tensor defined on $B$ captures the coupled effects of matter and gravitation.
When one projects this stress-energy tensor tangentially and normally to the spacelike 2-boundary of $B$, one obtains the quasilocal energy, momentum and spatial stress density.
The Brown-York quasilocal energy density (energy per 2-surface area) can be written as the following
\begin{equation}
\varepsilon=-\frac{1}{8\pi}\left(k-k_0\right),
\end{equation}
where $k$ is the extrinsic curvature of $\mathbb{S} $ when it is embedded in $\Sigma$ and $k_0$ is the corresponding extrinsic curvature evaluated for a suitable reference spacetime.
Starting with the action principle, one can show that the Brown-York quasilocal energy is the value of the 2-dimensional boundary Hamiltonian once the constraints are introduced in the bulk.
The Brown-York energy is given by
\begin{equation}
E_{BY}=-\frac{1}{8\pi}\oint _{\mathbb{S}}{d\mathbb{S}\left(k-k_0\right)}.
\end{equation}
\subsection{Kijowski (K) Energies}
Kijowski \cite{1997GReGr..29..307K} followed a Hamilton-Jacobi approach which is fundamentally different to that of Brown and York.
He attacked the quasilocal energy problem from first principles, by considering a symplectic structure on the phase space which is consistent with a nonvanishing boundary Hamiltonian\footnote{See \cite{Chen:1998aw} for a similar viewpoint.}.
This can be interpreted as a modification of the Arnowitt-Deser-Misner symplectic structure which assumes the vanishing of the boundary Hamiltonian under certain asymptotic conditions.
In fact, in collaboration with Tulczyjew, Kijowski used these symplectic structures to define boundary Hamiltonians \cite{Kijowski:1979, Kijowski:1984} decades before other quasilocal energy definitions were introduced.
This work appears to have been neglected by the community until Kijowski applied his approach to general relativity explicitly \cite{1997GReGr..29..307K}.

Kijowski applied Legendre transformations to the boundary Hamiltonian and obtained two physically meaningful quasilocal energy definitions.
He called them the \textit{internal energy} and \textit{free energy} due to them being related to each other via Legendre transformations with Dirichlet and Neumann type boundary conditions.
The two quasilocal energy definitions of Kijowski \cite{1997GReGr..29..307K} are respectively,
\begin{equation}\label{eq:E_Kin}
E_{K}^{in}=-\frac{1}{16\pi}\oint_{\mathbb{S}}{d\mathbb{S} \left[\frac{\left(k^2-l^2\right)-k_0^2}{k_0}\right]},
\end{equation}
\begin{equation}\label{eq:E_K}
E_{K}=-\frac{1}{8\pi}\oint_{\mathbb{S}}{d\mathbb{S} \left[\sqrt{\left(k^2-l^2\right)}-k_0\right]},
\end{equation}
where $k$ and $l$ are the trace of the extrinsic curvature of $\mathbb{S}$ with respect to its spacelike and timelike unit normals. Thus $\sqrt{\left(k^2-l^2\right)}$ gives twice the \textit{mean} extrinsic curvature of $\mathbb{S}$ \cite{Epp:2000zr}.

One may calculate $k_0$ by isometrically embedding $\mathbb{S}$ into Minkowski spacetime hyperplanes and dragging the 2-surface along the timelike Killing vector of the Minkowski spacetime.
The reference energy density, $k_0$, is the extrinsic curvature of the worldtube obtained in this manner and it is related to the boundary conditions.
Note that $E_{K}^{in}$ reduces to the Misner-Sharp-Hernandez energy for spherically symmetric spacetimes and gives the Hawking mass for the Schwarzschild geometry \cite{Szabados:2004vb}.
Therefore $E_{K}^{in}$ will be taken as the internal energy of the quasilocally defined system in this paper.
However, we do not follow Kijowski in referring to $E_{K}$ as the free energy.
This is because it takes the same form as the Liu-Yau quasilocal energy \cite{Liu:2003bx} which is interpreted differently in the literature, as we will now discuss.
\subsection{Liu-Yau (LY) Energy}
In Liu and Yau's work \cite{Liu:2003bx} there is no reference to a timelike 3-dimensional boundary.
They considered the embedding of $\mathbb{S}$ directly into a spacetime domain $D$ by taking its two normal null vectors and the corresponding mean extrinsic curvature.
That provides a well-defined quasilocal energy under the direct embedding of a 2-dimensional spacelike surface into a 4-dimensional spacetime.
This is the motivation for the method we use.
When Liu and Yau converted from their original notation into the one used here, their energy expression becomes:
\begin{equation}\label{eq:E_KLY}
E_{LY}=-\frac{1}{8\pi}\oint_{\mathbb{S}}{d\mathbb{S}\left[\sqrt{\left(k^2-l^2\right)}-k_0\right]}.
\end{equation}
The reference energy is obtained by embedding $\mathbb{S}$ into the 3-dimensional Euclidean space, $\mathbb{R}^3$, and calculating its extrinsic curvature, $k_0$.
This isometric embedding is unique up to the isometries of $\mathbb{R}^3$.
Note that their quasilocal energy expression is equal to Kijowski's energy given by (\ref{eq:E_K}).
The positivity of the Kijowski-Liu-Yau energy, denoted $E_{KLY}$ from now on, has been proven by Liu and Yau \cite{Liu:2003bx, 1081.83008}.
It is a widely accepted\footnote{Note that one can find a topologically spherical 2-surface in flat spacetime with negative Brown-York energy and positive Kijowski and Kijowski-Liu-Yau energies.
According to \'O Murchadha, Szabados and Tod \cite{O'Murchadha:2003xh} this might take place if $\mathbb{S}$ does not lie in a spacelike hyperplane of flat spacetime.
However, in the present paper we only consider spherical symmetry, in which case the Kijowski and Kijowski-Liu-Yau energies are well defined.
} quasilocal energy definition \cite{Murchadha:2007em,Szabados:2004vb}.

\subsection{On radial boost invariance}

The quasilocal energies $E_K^{in}$ and $E_{KLY}$ of a system are obtained via the mean extrinsic curvature of a 2-dimensional spacelike boundary, which we will call the \textit{screen}.
For a spherically symmetric spacetime, for example, the generator of the quasilocal energy is not just a single timelike vector.
Rather, one needs to consider both the future pointing timelike normal and the outward pointing radial spacelike normal in order to calculate the mean extrinsic curvature of the screen.

Following the ideas of Epp\footnote{Epp defined a similar energy to $E_{KLY}$ with a different nonunique choice of reference energy.}
\cite{Epp:2000zr} $E_{KLY}$ can be interpreted as a \textit{proper} mass-energy of the system. It is known that both $E_{KLY}$ and $E_{K}^{in}$ are invariant under radial boosts of the quasilocal observers who define the 2-surface \cite{1997GReGr..29..307K,Epp:2000zr,Liu:2003bx,Szabados:2004vb}.
Such a property
 is necessary to define a system consistently since one needs the ability to keep constant the degrees of freedom that \textit{define} the screen enclosing the system.
In the case here, $\{ \theta, \phi \}$ are the coordinates on the screen that are kept constant.
Then the evolution of the system is investigated by perturbing the screen along the remaining degrees of freedom, parametrized by the $\{ t, r \}$ coordinates.
Thus the screen observers agree on the quasilocal energy content of the \textit{same} system irrespective of them being boosted or having instantaneous radial accelerations with respect to any other screen.

We note that in this picture, the Kodama vector is an object that lives in the temporal-radial plane.
The fact that the Kodama vector is associated with a conserved Misner-Sharp-Hernandez energy has previously been described as a \textit{miracle} \cite{PhysRevD.82.044027}.
Here we emphasize that since $E_K^{in}$ matches $E_{MSH}$ under spherical symmetry, this miracle is a natural consequence of any consistent quasilocal Hamiltonian formalism of general relativity.
Thus one does not need to define a single \textit{preferred} observer in the energy calculations.

The distinction of the degrees of freedom that are used to define the system and the ones used in the investigation of the evolution is crucial for the interpretation of the formalism introduced in the next section.

\section{Geometry of worldsheet focusing and gravitational thermodynamics}\label{geoandequil}
Capovilla and Guven \cite{Capovilla:1994yk} generalize the Raychaudhuri equation which gives the focusing of an arbitrary dimensional timelike worldsheet that is embedded in an arbitrary dimensional spacetime.
We apply their formalism to a 2-dimensional timelike worldsheet, $\mathbb{T}$, embedded in a 4-dimensional spacetime, using their original notation.

Consider an embedding of an oriented worldsheet with an induced metric, $\tensor{\eta}{_a_b}$, written in terms of orthonormal basis tangent vectors, $\{ \tensor{E}{_a} \}$,
\begin{equation}
g({\tensor{E}{_a}},{\tensor{E}{_b}})={\tensor{\eta}{_a_b}},
\end{equation}
where $g_{\mu \nu}$ is the 4-dimensional spacetime metric.
Now consider the two unit normal vectors, $\{ \tensor{n}{^i} \}$, of the worldsheet which are defined up to a local rotation by,
\begin{equation}
g({\tensor{n}{_i}},{\tensor{n}{_j}})={\tensor{\delta}{_i_j}}
\end{equation}
\begin{equation}
g({\tensor{n}{^i}},{\tensor{E}{_a}})=0,
\end{equation}
where $\{a,b\}=\{\hat{0},\hat{1}\}$ and $\{i,j\}=\{\hat{2},\hat{3}\}$ are the diad indices and the Greek indices refer to 4-dimensional spacetime coordinates.
Also note that to raise (or lower) the indices of tangential and normal diad indices one should use ${\tensor{\eta}{^a^b}}$ (or ${\tensor{\eta}{_a_b}}$) and ${\tensor{\delta}{^i^j}}$ (or ${\tensor{\delta}{_i_j}}$) respectively.

At this point it is important to distinguish the different covariant derivative operators \cite{Capovilla:1994yk}.
Let the torsionless covariant derivatives defined by the spacetime coordinate metric be ${\tensor{D}{_\mu}}$ and their projection onto the worldsheet be denoted by ${\tensor{D}{_a}}={\tensor{E}{^\mu_a}}{\tensor{D}{_\mu}}$.
On the worldsheet $\mathbb{T}$, we introduce two covariant derivatives. ${\tensor{\nabla}{_a}}$ is defined with respect to the intrinsic metric and ${\tensor{\tilde{\nabla}}{_a}}$ is defined on tensors under rotations of the normal frame on $\mathbb{S}$.
To study the deformations of $\mathbb{T}$ and $\mathbb{S}$, the following extrinsic variables are introduced \cite{Capovilla:1994yk}.
The extrinsic curvature, Ricci rotation coefficients and extrinsic twist of
$\mathbb{T}$ are respectively defined by,
\begin{equation}
\tensor{K}{_{ab}^i}=-\tensor{g}{_\mu _\nu}\left(\tensor{D}{_a}\tensor{E}{^\mu _b}\right) \tensor{n}{^\nu ^i}=\tensor{K}{_{ba}^i},
\end{equation}
\begin{equation}
\tensor{\gamma}{_a_b_c}=\tensor{g}{_\mu _\nu}\left(\tensor{D}{_a}\tensor{E}{^\mu _b}\right) \tensor{E}{^\nu _c}=-\tensor{\gamma}{_a_c_b},
\end{equation}
\begin{equation}
\tensor{w}{_a^i^j}=\tensor{g}{_\mu _\nu}\left(\tensor{D}{_a}\tensor{n}{^\mu ^i}\right) \tensor{n}{^\nu^j}=-\tensor{w}{_a^j^i}
\end{equation}
while the extrinsic curvature, Ricci rotation coefficients and extrinsic twist of
$\mathbb{S}$ are respectively defined by,
\begin{equation}
\tensor{J}{_a^i^j}=\tensor{g}{_\mu _\nu}\left(\tensor{D}{^i}\tensor{E}{^\mu _a}\right) \tensor{n}{^\nu ^j},
\end{equation}
\begin{equation}
\tensor{\gamma}{_i_j_k}=\tensor{g}{_\mu _\nu}\left(\tensor{D}{_i}\tensor{n}{^\mu _j}\right) \tensor{n}{^\nu _k}=-\tensor{\gamma}{_i_k_j},
\end{equation}
\begin{equation}
\tensor{S}{_a_b^i}=\tensor{g}{_\mu _\nu}\left(\tensor{D}{^i}\tensor{E}{^\mu _a}\right) \tensor{E}{^\nu _b}=-\tensor{S}{_b_a^i}.
\end{equation}
By using those extrinsic variables one can investigate how the orthonormal basis $\{ \tensor{E}{_a},\tensor{n}{^i} \}$ varies when perturbed on $\mathbb{T}$ according to,
\begin{equation}
\tensor{D}{_a}\tensor{E}{_b}=\tensor{\gamma}{_a_b^c}\tensor{E}{_c}-\tensor{K}{_a_b^i}\tensor{n}{_i},
\end{equation}
\begin{equation}
\tensor{D}{_a}\tensor{n}{^i}=\tensor{K}{_a_b^i}\tensor{E}{^b}+\tensor{w}{_a^i^j}\tensor{n}{_j},
\end{equation}
or perturbed on $\mathbb{S}$ according to,
\begin{equation}
\tensor{D}{_i}\tensor{E}{_a}=\tensor{S}{_a_b_i}\tensor{E}{^b}+\tensor{J}{_a_i_j}\tensor{n}{^j},
\end{equation}
\begin{equation}
\tensor{D}{_i}\tensor{n}{_j}=-\tensor{J}{_a_i_j}\tensor{E}{^a}+\tensor{\gamma}{_i_j^k}\tensor{n}{_k}.
\end{equation}
Then the generalized Raychaudhuri equation which quantifies how much a worldsheet focuses is found to be \cite{Capovilla:1994yk}
\begin{eqnarray}\label{Raychaudhuri}
\tensor{\tilde{\nabla}}{_b}\tensor{J}{_a^i^j} &=& -\tensor{\tilde{\nabla}}{^i}\tensor{K}{_a_b^j}-\tensor{J}{_b^i_k}\tensor{J}{_a^k^j}-\tensor{K}{_b_c^i}\tensor{K}{_a^c^j} +g\left(R(\tensor{E}{_b},\tensor{n}{^i})\tensor{E}{_a},\tensor{n}{^j}\right),
\end{eqnarray}
where $\tensor{R}{^\alpha_\beta_\mu_\nu}$ is the Riemann tensor of the 4-dimensional spacetime \cite{Capovilla:1994yk}, and
\begin{equation}
g(R(\tensor{E}{_a},\tensor{n}{_i})\tensor{E}{_b},\tensor{n}{_j})=\tensor{R}{_\alpha_\beta_\mu_\nu}\tensor{E}{^\mu_a}\tensor{n}{^\nu_i}\tensor{E}{^\beta_b}\tensor{n}{^\alpha_j}.
\label{Rproj}\end{equation}
Note that $\tensor{w}{_b_i^k}$ transforms as a connection under the rotation of $\mathbb{S}$ and
\begin{equation}\label{eq:CurlyCovJ}
\tensor{\tilde{\nabla}}{_b}\tensor{J}{_a_i_j}=\underbrace{\tensor{\nabla}{_b}\tensor{J}{_a_i_j}}_\text{$\tensor{D}{_b}\tensor{J}{_a_i_j}-\tensor{\gamma}{_b_a^c}\tensor{J}{_c_i_j}$}-\tensor{w}{_b_j^k}\tensor{J}{_a_k_i}-\tensor{w}{_b_i^k}\tensor{J}{_a_k_j}.
\end{equation}
Likewise, $\tensor{S}{_a_b^i}$ transforms as a connection under the rotation of $\mathbb{T}$ such that
\begin{equation}\label{eq:CurlyCovK}
\tensor{\tilde{\nabla}}{_i}\tensor{K}{_a_b^j}=\underbrace{\tensor{\nabla}{_i}\tensor{K}{_a_b^j}}_\text{$\tensor{D}{_i}\tensor{K}{_a_b^j}-\tensor{\gamma}{_i^j_k}\tensor{K}{_a_b^k}$}-\tensor{S}{_a_c_i}\tensor{K}{_b^c^j}-\tensor{S}{_b_c_i}\tensor{K}{_a^c^j}.
\end{equation}
If one contracts (\ref{Raychaudhuri}) with the orthogonal basis metrics
${\tensor{\eta}{^a^b}}$ and ${\tensor{\delta}{_i_j}}$ one finds
\begin{eqnarray}\label{eq:ContCurlyCovJ}
\left(\tensor{\tilde{\nabla}}{_b}\tensor{J}{_a^i^j}\right){\tensor{\eta}{^a^b}}\tensor{\delta}{_i_j} &=&-\left(\tensor{\tilde{\nabla}}{^i}\tensor{K}{_a_b^j}\right){\tensor{\eta}{^a^b}}\tensor{\delta}{_i_j}-\tensor{J}{_b^i_k}\tensor{J}{_a^k^j}{\tensor{\eta}{^a^b}}\tensor{\delta}{_i_j} -\tensor{K}{_b_c^i}\tensor{K}{_a^c^j}{\tensor{\eta}{^a^b}}\tensor{\delta}{_i_j}\nonumber \\
&&+g(R(\tensor{E}{_b},\tensor{n}{^i})\tensor{E}{_a},\tensor{n}{^j}){\tensor{\eta}{^a^b}}\tensor{\delta}{_i_j},
\end{eqnarray}
which is the equation to which we will assign a quasilocal thermodynamic
interpretation to arrive at the key results of this paper.

In the 3+1 formalism, the Raychaudhuri equation \cite{1955PhRv...98.1123R} for a congruence of timelike worldlines tells us how much the congruence expands or contracts.
For a timelike congruence, that is not necessarily geodesic, with vanishing shear and vorticity, the Raychaudhuri equation reads \cite{1984ucp..book.....W}
\begin{equation}
-\frac{d\Theta}{d\lambda}=+\frac{1}{3}\Theta^2+R_{\mu \nu}v^{\mu}v^{\nu}-D _{\mu}\left(v^{\nu}D _{\nu}v^{\mu}\right),
\end{equation}
where $\Theta$ is the expansion scalar, $\lambda$ is the affine parameter, $R_{\mu \nu}$ is the Ricci tensor and $v^{\mu}$ are the components of the 4-velocity vector field of an observer.
In the 2+2 language, however, the generalized Raychaudhuri equation (\ref{eq:ContCurlyCovJ}) tells us how much a congruence of worldsheets, rather than the worldlines, focuses.

Now consider a general spherically symmetric spacetime.
For radially moving observers, the extrinsic curvature and the extrinsic twist of $\mathbb{T}$ vanishes as well as the extrinsic twist of $\mathbb{S}$.
Then (\ref{eq:ContCurlyCovJ}) reduces to
\begin{eqnarray}\label{eq:ContCurlyCovJsph}
-\left(\tensor{\tilde{\nabla}}{_b}\tensor{J}{_a^i^j}\right){\tensor{\eta}{^a^b}}\tensor{\delta}{_i_j}&=&\tensor{J}{_b^i_k}\tensor{J}{_a^k^j}{\tensor{\eta}{^a^b}}\tensor{\delta}{_i_j}-g(R(\tensor{E}{_b},\tensor{n}{^i})\tensor{E}{_a},\tensor{n}{^j}){\tensor{\eta}{^a^b}}\tensor{\delta}{_i_j}.
\end{eqnarray}
We will interpret (\ref{eq:ContCurlyCovJsph}) as a thermodynamic relation for a quasilocally defined system. Note that the choice of the orthonormal dyad metrics are $\eta _{ab}=\left(-1,1\right)$ and $\delta _{ij}=\left(1,1\right)$ for the rest of the paper.
\subsection{Quasilocal thermodynamic equilibrium conditions}

At this point one needs to take care with the definitions of the thermodynamic variables under equilibrium and nonequilibrium conditions.
In general, equilibrium can be seen as a specific state of a system that is ordinarily in nonequilibrium with its surroundings.
Given specific equilibrium conditions, the equilibrium state acts as an \textit{attractor} to bring the system into a preferably stable state \cite{1978Sci...201..777P}.
Moreover, in classical thermodynamics, only in the equilibrium case are the existence of thermodynamic potentials guaranteed \cite{1984PhRvL..52....9G}.
For such equilibrium states the thermodynamic potentials are Lyapunov functions\footnote{
Lyapunov functions are nonnegative functions that have at least one local maxima or minima at a point of interest. They are continuous functions with continuous first order derivatives and they vary monotonically with the evolution parameter \cite{stratonovich1994nonlinear}.},
and they can be written as \textit{linear} combinations of each other \cite{Demirel2007xix}.

Let us recall the definitions of thermodynamic potentials in classical thermodynamics at the equilibrium state \cite{1983eqth.book.....A},
\begin{eqnarray}
Helmholtz\ Free \ Energy:\FF\doteq \mathpzc{U}-\mathpzc{T}\,\mathpzc{S}, \label{eq:HelmFree}\\
Gibbs\ Free\ Energy: \mathpzc{G}\doteq \FF+\mathpzc{W}, \label{eq:GibssFree}\\
Enthalpy: \mathpzc{H}\doteq \mathpzc{G}+\mathpzc{T}\,\mathpzc{S}\doteq \mathpzc{U}+\mathpzc{W}, \label{eq:Enthalpy}
\end{eqnarray}
where $\mathpzc{U}$ is the internal energy, and $\mathpzc{W}$ represents the work terms which may include $PV$ type and other types of work in general. The `$\doteq$' sign will be used for equations that hold \textit{only} at quasilocal thermodynamic equilibrium from now on.

Note that Helmholtz free energy is the amount of reversible work done on a system in an isothermal process \cite{1969stph.book.....L}.
It is one of the thermodynamic variables that can be defined both in equilibrium and in nonequilibrium states \cite{attard2002thermodynamics}.
Moreover, one way of defining the thermodynamic equilibrium is to set the Helmholtz free energy to its minimum value \cite{Demirel2007xix}.
For this reason the Helmholtz free energy provides a physically natural means to interpret (\ref{eq:ContCurlyCovJsph}) thermodynamically.
\subsubsection{Helmholtz free energy density}
We will take the extrinsic curvature of $\mathbb{S}$ as a measure of the matter + gravitational Helmholtz free energy density and define\footnote{The $\tensor{J}{_a^i^j}\tensor{J}{^a_j_i}/4$ term appears in the definition of the Hawking \cite{Hawking:1968qt} and Liu-Yau \cite{Liu:2003bx} mass-energies, as the term $\mu \rho$ in the notation of these authors.}
\begin{equation}\label{eq:fsquare}
\mathpzc{f} ^a \mathpzc{f} _a \equiv 2 \left(\tensor{J}{_a^i^j}\tensor{J}{^a_j_i}\right).
\end{equation}
since
\begin{equation}\label{eq:fdensity}
\mathpzc{f} \equiv \sqrt{\mathpzc{f}^a\mathpzc{f}_a}=\sqrt{k^2-l^2}.
\end{equation}
is the object that appears in the quasilocal energy definitions of section~\ref{energydef}.
Then the Helmholtz free energy of the system is obtained once $\mathpzc{f}$ is integrated on $\mathbb{S}$, i.e.,
\begin{equation}\label{eq:Helmholtz}
\FF=\frac{1}{16\pi}\oint_{\mathbb{S}}{\mathpzc{f}\cdot d \mathbb{S}}.
\end{equation}
Since the equilibrium condition is defined in this case by the minimum of the Helmholtz free energy, other thermodynamic potentials should be written as linear combinations of each other once one sets $\FF=\FF_{min}$.
Thus at equilibrium, the Gibbs free energy and the internal energy should read
\begin{equation}\label{eq:withFmin}
\mathpzc{G}\doteq \FF_{min}+\mathpzc{W},
\end{equation}
\begin{equation}
\mathpzc{U}\doteq \FF_{min}+\mathpzc{T}\,\mathpzc{S}.
\end{equation}
where $\mathpzc{W}$, $\mathpzc{T}$ and $\mathpzc{S}$ are to be defined.
The Helmholtz free energy density defined by (\ref{eq:fsquare})and (\ref{eq:fdensity}) is required
to be a nonnegative real scalar, and the minimum value it can take is zero.
This brings us to write the equations above with $\FF_{min}=0$ as
\begin{equation}\label{eq:GwithFmin}
\mathpzc{G}\doteq \mathpzc{W},
\end{equation}
\begin{equation}\label{eq:UwithFmin}
\mathpzc{U}\doteq \mathpzc{T}\,\mathpzc{S}.
\end{equation}
Then when $\mathpzc{f}=\sqrt{\mathpzc{f}^a\mathpzc{f}_a}=\sqrt{k^2-l^2}\doteq 0$, (\ref{eq:ContCurlyCovJsph}) becomes
\begin{equation} \label{eq:ContCurlyCovJspheq}
-\tensor{\tilde{\nabla}}{_a}\tensor{J}{^a^i_i}\doteq -g(R(\tensor{E}{_b},\tensor{n}{^j})\tensor{E}{^b},\tensor{n}{_j}).
\end{equation}
Recalling the fact that thermodynamic potentials are nonnegative real functions at equilibrium, we will force the terms in (\ref{eq:ContCurlyCovJspheq}) to take nonnegative values by taking the absolute value of each side before we start making further quasilocal thermodynamic interpretations. Thus,
\begin{equation} \label{eq:ContCurlyCovJspheqabs}
\left|-\tensor{\tilde{\nabla}}{_a}\tensor{J}{^a^i_i}\right|\doteq \left|-g(R(\tensor{E}{_b},\tensor{n}{^j})\tensor{E}{^b},\tensor{n}{_j})\right|.
\end{equation}
\subsubsection{Work density}
We will now give a thermodynamic interpretation to the quantity on the r.h.s. of (\ref{eq:ContCurlyCovJspheqabs}).
In the 3+1 formalism, when one considers two observers on neighbouring timelike geodesics the deviation of the geodesics determines the relative accelerations of the observers.
If we consider the spacelike separation 4-vector, $\vec{\xi}$, that connects the neighbouring geodesics, then the components of the relative tidal acceleration are given by \cite{1984ucp..book.....W}
\begin{equation}\label{eq:relacc}
\frac{d^2\xi ^\mu}{d\tau ^2}=\tensor{R}{^\mu_\nu _\rho _\sigma}v^{\nu}v^{\rho}\xi ^{\sigma},
\end{equation}
where $\tau$ is the proper time.
Thus for a spherically symmetric spacetime we define a \textit{relative work} term that mimics $W=\vec{F}\cdot \vec{x}$ by
\begin{equation}\label{eq:relwork}
\left(\frac{d^2\xi ^\mu}{d\tau ^2}\right)\xi_{\mu}=\tensor{R}{^\gamma _\nu _\rho _\sigma}v^{\nu}v^{\rho}\xi^{\sigma}\xi_{\gamma}.
\end{equation}
This relative work term can be interpreted as a measure of energy expended within the surface of a body to stretch or contract it under the influence of tidal forces, if we assume $\vec{\xi}$ lives on the screen, $\mathbb{S}$.

Our interpretation is similar to that of Schutz \cite{1985gasr.book..237S} who considered the limits of validity of the geodesic deviation equation and calculated the second order contributions.
He also acknowledged the fact that connecting two geodesics with a separation vector is essentially nonlocal.
Thus the reason (\ref{eq:relacc}) is valid only for nearly parallel, neighbouring geodesics is simply due to observers trying to measure a nonlocal quantity, locally.
Consequently (\ref{eq:relwork}) has a more fundamental quasilocal interpretation and the $\left|-g(R(\tensor{E}{_b},\tensor{n}{^j})\tensor{E}{^b},\tensor{n}{_j})\right|$ term on the r.h.s. of (\ref{eq:ContCurlyCovJspheqabs}) might be taken as a measure of {\em work density} attributed to $\mathbb{S}$.

To understand this intuitively, consider the analogy of a soap bubble.
The work done per area to create the surface of a bubble is \cite{1983eqth.book.....A}
\begin{equation}\label{eq:surftension}
\mathpzc{W}_{class}=\oint {\gamma \cdot dA},
\end{equation}
where $\gamma$ is the surface tension and $dA$ is a surface area element of the bubble.
According to classical theory, surface tension arises due to the unbalanced intermolecular forces in the bubble. Likewise, according to the analogy formed here, $\left|-g(R(\tensor{E}{_b},\tensor{n}{^j})\tensor{E}{^b},\tensor{n}{_j})\right|$ is a measure of energy density due to the relative tidal forces that observers experience when they move along radial worldlines.
This is of course applicable for observers who share the same screen $\mathbb{S}$.
Therefore at quasilocal thermodynamic equilibrium, we can define a general work density, namely a type of surface tension, according to
\begin{equation}\label{eq:wdensity}
\mathpzc{w}\equiv \sqrt{\mathpzc{w}^a \mathpzc{w}_a}\equiv \sqrt{2\left|-g(R(\tensor{E}{_b},\tensor{n}{^j})\tensor{E}{^b},\tensor{n}{_j})\right|},
\end{equation}
so that the amount of corresponding work is given by
\begin{equation}\label{eq:work}
\mathpzc{W}\doteq \frac{1}{16\pi}\oint_{\mathbb{S}}{\mathpzc{w} \cdot d \mathbb{S}}.
\end{equation}
\subsubsection{Gibbs free energy density}
In classical thermodynamics, when equilibrium is defined by the minimum of the Helmholtz free energy, the Gibbs free energy reads \cite{1983eqth.book.....A}
\begin{equation}
\mathpzc{G}_{class}\doteq \oint {\gamma \cdot dA},
\end{equation}
for the thermodynamics of surfaces with constant pressure.
Following the analogy with the surface of a soap bubble,
\begin{equation}
\mathpzc{W}_{class}=\oint {\gamma \cdot dA} \Leftrightarrow \mathpzc{W} \doteq \oint_{\mathbb{S}}{\mathpzc{w} \cdot d \mathbb{S}},
\end{equation}
so that
\begin{equation}
\mathpzc{G}_{class}\doteq \oint {\gamma \cdot dA} \Leftrightarrow \mathpzc{G}\doteq \mathpzc{W}
\end{equation}
should hold.
This is consistent with (\ref{eq:GwithFmin}) which states that $\mathpzc{G}\doteq \mathpzc{W}$ since the minimum Helmholtz free energy is zero according to the equilibrium condition defined here.
Thus, the l.h.s.\ of (\ref{eq:ContCurlyCovJspheqabs}) can be taken as a measure of the quasilocal Gibbs free energy density:
\begin{equation}\label{eq:gibbsdensity}
\mathpzc{g}\equiv \sqrt{\mathpzc{g}^a\mathpzc{g}_a}\equiv \sqrt{2\left|-\tensor{\tilde{\nabla}}{_a}\tensor{J}{^a^i_i}\right|},
\end{equation}
from which the Gibbs energy can be obtained by
\begin{equation}\label{gibbsenergy}
\mathpzc{G}\doteq \frac{1}{16\pi}\oint_{\mathbb{S}}{\mathpzc{g}\cdot d \mathbb{S}}.
\end{equation}

Note that, in general, the Raychaudhuri equation becomes nonlinear if one wants to write it in terms of the energy densities defined here.
However, recall that the existence of thermodynamic potentials is guaranteed only in the equilibrium case in which the potentials can be written linearly in terms of each other.
In classical surface thermodynamics, the surface tension, pressure gradient across the surface and mean curvature of the surface can be related via the Young-Laplace equation \cite{Rusanov2005111}.
Ideally, in order to reach the equilibrium, fluids tend to minimize their surfaces until they have zero mean curvature.
This is when the the surface tension takes its critical value.
In the formalism presented here, which is in line with our analogy, this happens when $\mathpzc{f}=\sqrt{\mathpzc{f}^a \mathpzc{f}_a}=\sqrt{k^2-l^2}\doteq 0$, which defines the apparent horizon of a given spacetime.

Here we use a general apparent horizon \cite{Hayward:1993wb}, defined by the marginal surfaces on which at least one of the expansion scalars of the null congruences is zero, i.e., $\theta _{(l)}\theta _{(n)}=0$, where $l^a$ ($n^a$) is the outward (inward) pointing future-directed normal. Both the conditions $\{ \theta _{(l)}>0 ,\theta _{(n)}=0 \}$ and $\{ \theta _{(l)}=0 ,\theta _{(n)}<0 \}$ have previously been used to define apparent horizons \cite{Faraoni:2011hf,Booth:2005qc,Tian:2014sca}. Here $\tensor{J}{_a^i^j}\tensor{J}{^a_j_i}$ in (\ref{eq:fsquare}) gives a measure of $\theta _{(l)}\theta _{(n)}$. Thus when it is equated to zero, one can conclude that at least one of the expansion scalars of the incoming or outgoing null congruences converges without knowing which one actually does.

\subsubsection{Internal energy density}
On introducing the quasilocal energy definitions in section~\ref{energydef}, we stated that $E_K^{in}$ is a good candidate for the total matter + gravitational energy content of a system.
It is derived via a Hamilton-Jacobi formalism with Dirichlet boundary conditions.
According to Kijowski those boundary conditions are associated with the true degrees of freedom of the quasilocally defined domain that gives the \textit{true energy} \cite{1997GReGr..29..307K}.
When the equilibrium condition is imposed, the internal energy density in (\ref{eq:E_Kin}) can be written as
\begin{equation}
\mathpzc{u} \doteq k_0.
\end{equation}
Thus the quasilocal internal energy at equilibrium becomes
\begin{equation}\label{eq:internalateq}
\mathpzc{U} \doteq \frac{1}{16\pi}\oint _{\mathbb{S}}{k_0}\cdot d\mathbb{S}.
\end{equation}
which should satisfy the equilibrium condition (\ref{eq:UwithFmin}) without any
$P V$ type term\footnote{When Kijowski applied the boundary conditions in his original derivation \cite{1997GReGr..29..307K}, he chose the induced metric components of $\mathbb{S}$ to be time independent.
This choice holds both for $E_{K}^{in}$ and for $E_{KLY}$ and it is set to prevent the extra inclusion of \textit{volume acquisitions} in the energy definitions.
However, Liu and Yau do not set such a condition on $E_{KLY}$.
Moreover, according to the equilibrium conditions set here, $\mathpzc{U}$ contains only the $\mathpzc{T}\,\mathpzc{S}$ term. Therefore in our case $E_{K}^{in}$ can be safely used for generic coordinate representations of spherically symmetric spacetimes at equilibrium.}.
This requires that we define a quasilocal entropy and temperature at
equilibrium.

Since the 2-surface $\mathbb{S}$ located at the generalized apparent horizon
enters naturally, we can follow the traditional approach of Bardeen
\cite{PhysRevD.7.2333} and define the quasilocal equilibrium entropy:
\begin{equation}\label{eq:entropy}
\mathpzc{S} \doteq \frac{Area\left(\mathbb{S}\right)}{\lambda},
\end{equation}
where $\lambda$ is a constant which is usually taken to be $4$ in
gravitational equilibrium thermodynamics of horizons.
By (\ref{eq:UwithFmin}), at equilibrium the quasilocal temperature of the
system is then given by
\begin{equation}\label{eq:temperature}
\mathpzc{T}\doteq \frac{\mathpzc{U}}{\mathpzc{S}}\,
\end{equation}
{\em without any direct reference to surface gravity.}
\subsubsection{First law of thermodynamics}
According to the formalism constructed here, the first law should be written as
\begin{equation}
\delta \mathpzc{U}\doteq \delta \left(\mathpzc{T}\,\mathpzc{S}\right)\doteq \mathpzc{T} \, \delta \mathpzc{S},
\end{equation}
since we defined the quasilocal thermodynamic equilibrium via the minimization of the Helmholtz free energy which is applicable for isothermal processes.
The problem with some of the gravitational thermodynamic constructions is that the total variation of the internal energy and entropy is performed in specific directions for which the first law does not have the dimensions of energy\footnote{For example see \cite{Faraoni:2011hf}.}.
However, in our framework the quasilocal behaviour of the system sets the degrees of freedom with respect to which the total variation can be defined.
These are the degrees of freedom that live on the instantaneously defined timelike surface $\mathbb{T}$.
Hence, we will set the total variation to be the one on the worldsheet and write
\begin{equation}
\delta \mathpzc{U}\equiv \frac{1}{2}\sqrt{\tilde{\nabla}_a\mathpzc{U}\,\tilde{\nabla}^a\mathpzc{U}}.
\end{equation}
Thus the first law should read
\begin{equation}\label{eq:firstlaw}
\frac{1}{2}\sqrt{\tilde{\nabla}_a\mathpzc{U}\,\tilde{\nabla}^a\mathpzc{U}}\doteq \mathpzc{T}\left(\frac{1}{2}\sqrt{\tilde{\nabla}_a\mathpzc{S}\,\tilde{\nabla}^a\mathpzc{S}}\right)
,\end{equation}
with
\begin{equation}\label{eq:constTemp}
\delta \mathpzc{T}\equiv \frac{1}{2}\sqrt{\tilde{\nabla}_a\mathpzc{T}\,\tilde{\nabla}^a\mathpzc{T}}\doteq 0,
\end{equation}
where $\{a,b\} = \{\hat{0},\hat{1}\}$.
Thus the temperature will be a \textit{worldsheet--constant} rather than a constant with respect to some coordinate time.
For the examples that are presented in the next section, it is easy to check that (\ref{eq:constTemp}) is satisfied.
Note that, in general,
\begin{equation}
\tilde{\nabla}_a \rightarrow {\nabla}_a \rightarrow D_a \rightarrow \tensor{E}{^\mu _a}D_{\mu} \rightarrow \tensor{E}{^\mu _a} \partial _{\mu},
\end{equation}
can be used for the scalar functions that appear in (\ref{eq:firstlaw}) and (\ref{eq:constTemp}).

\section{Examples}\label{examples}
\subsection{Schwarzschild geometry}
Consider the Schwarzschild metric in standard coordinates
\begin{equation}\label{eq:Schmetric}
ds^2=-\left(1-\frac{2M}{r}\right)dt^2+\left(1-\frac{2M}{r}\right)^{-1}dr^2+r^2d \Omega^2,
\end{equation}
where $d\Omega^2=d\theta ^2+\sin ^2 \theta d\phi ^2$.
As stated previously, to define a consistent system observers should have fixed angular coordinates.
As one example consider static radial observers with double diad
\begin{eqnarray}
\tensor{E}{^\mu _{\hat{0}}}=\left(\frac{1}{\sqrt{\left(1-\frac{2M}{r}\right)}},0,0,0\right), \qquad \, \, \, \tensor{n}{^\mu _{\hat{2}}}=\left(0,0,\frac{1}{r},0\right),\\
\tensor{E}{^\mu _{\hat{1}}}=\left(0,\sqrt{\left(1-\frac{2M}{r}\right)},0,0\right), \qquad \tensor{n}{^\mu _{\hat{3}}}=\left(0,0,0,\frac{1}{r\sin \theta}\right).
\end{eqnarray}
The choice of static observers is inconsequential for our results, which also
apply to observers with an arbitrary instantaneous radial boost with respect
to this frame. For such observers, (\ref{eq:fsquare}) implies
\begin{equation}
\mathpzc{f}=\sqrt{\mathpzc{f}_a \mathpzc{f}^a}=2\sqrt{\left(\frac{r-2M}{r^3}\right)},
\end{equation}
which can be substituted in the Raychaudhuri equation,~(\ref{eq:ContCurlyCovJsph}), in the general nonequilibrium case to give a notion of nonequilibrium quasilocal energy exchange.

In order to set the quasilocal thermodynamic equilibrium condition, $\FF$ should be minimized.
It is easy to see that, this occurs when $r=2M$ which coincides with the location of the black hole horizon.
Now let us calculate the internal energy at equilibrium.
Given the metric in (\ref{eq:Schmetric}) and the isometric embedding of $\mathbb{S}$ into Euclidean 3-space, one finds $k_0=2/r$.
Then according to (\ref{eq:UwithFmin}) and (\ref{eq:internalateq}),
\begin{equation}
\mathpzc{U}\doteq \mathpzc{T}\,\mathpzc{S} \doteq M,
\end{equation}
with $\mathpzc{T}\doteq \lambda/(8\pi r) \doteq\lambda/(16\pi M)$,
and $\mathpzc{S}\doteq Area(\mathbb{S})/\lambda \doteq (16 \pi M^2)/{\lambda}$ where $\lambda$ is a constant.
For those who wish to relate this temperature to the Hawking temperature \cite{hawking1975}, there is a problem of factor of two, which has been encountered in similar context before \cite{'tHooft:1983ap,'tHooft:1984re,Akhmedov:2006pg,Nakamura:2007vp,Pilling:2007cn} . For $\lambda=4$ the temperature gives twice the Hawking temperature, i.e., $\mathpzc{T}=2T_H=2\left(8\pi M\right)^{-1}$.
The literature is divided into two camps when it comes to the value of the temperature of radiation for a particle that tunnels through the horizon.
Usually, those who favour $\mathpzc{T}=2T_H$ also favour the idea of dividing the entropy by 2 in order to satisfy Hawking's original first law \cite{Akhmedov:2006pg,Pilling:2007cn}.
However, according to Hawking's original first law \cite{hawking1975}, for a static black hole \cite{0034-4885-41-8-004},
\begin{equation}\label{orgnlfirstlaw}
\frac{Energy}{2}=Temperature \times Entropy.
\end{equation}
Thus, if $\lambda=4$ then one should \textit{not} divide the original entropy expression by 2 in order to get the correct internal energy on the l.h.s. of (\ref{orgnlfirstlaw}).

Also it is easy to check (\ref{eq:constTemp}),
\begin{equation}
\delta \mathpzc{T}=\frac{1}{2}\sqrt{\tilde{\nabla}_a \left(\frac{\lambda}{8\pi r}\right)\tilde{\nabla}^a \left(\frac{\lambda}{8\pi r}\right)}\doteq 0.
\end{equation}
This states that the system can be assigned a single temperature value which is a worldsheet--constant.
One also finds the corresponding work (\ref{eq:work}) term and Gibbs free energy (\ref{gibbsenergy}) term
\begin{equation}
\label{eq:WorkSch}
\mathpzc{W}\doteq \frac{1}{16\pi}\sqrt{2\bigg|\frac{-4M}{r^3}\bigg|}\left(4\pi r^2\right), \qquad
\mathpzc{G} \doteq \frac{1}{16\pi}\sqrt{2\bigg|\frac{2(r-4M)}{r^3}\bigg|} \left(4\pi r^2\right)
\end{equation}
which are equal at the quasilocal thermodynamic equilibrium, i.e.,
\begin{equation}\label{eq:GibbsWorkSch}
\mathpzc{G} \doteq \mathpzc{W}\doteq M.
\end{equation}
\subsection{Friedmann-Lema\^{\i}tre-Robertson-Walker (FLRW) geometry}
Now consider the FLRW metric in comoving coordinates
\begin{equation}
ds^2=-dt^2+\left(\frac{a^2(t)}{1-\kappa r^2}\right)dr^2+a^2(t)r^2d \Omega^2,
\end{equation}
where $\kappa =\{-1,0,1\}$ for open, flat and closed universes respectively and
the comoving observer dyads are
\begin{eqnarray}
\tensor{E}{^\mu _{\hat{0}}}=\left(1,0,0,0\right), \qquad \qquad \qquad \, \tensor{n}{^\mu _{\hat{2}}}=\left(0,0,\frac{1}{ar},0\right),\\
\tensor{E}{^\mu _{\hat{1}}}=\left(0,\frac{\sqrt{1-\kappa r^2}}{a},0,0\right), \qquad \tensor{n}{^\mu _{\hat{3}}}=\left(0,0,0,\frac{1}{ar\sin \theta}\right).
\end{eqnarray}

Again note that the resultant thermodynamic potential densities do not change for observers with an arbitrary instantaneous radial boost with respect to the comoving observers.
For such a set up, Helmholtz free energy density is
\begin{equation}
\mathpzc{f}=\sqrt{\mathpzc{f}_a \mathpzc{f}^a}=\sqrt{2\left(\frac{2-2\kappa r^2-2\dot{a}^2 r^2}{a^2r^2}\right)}.
\end{equation}\\
If we consider the equilibrium case where free energy takes its minimum value, one can find the equilibrium condition to be
\begin{equation}\label{eq:FLRWeqcond}
r\doteq \frac{1}{\sqrt{\kappa +\dot{a}^2}}, \qquad or \qquad \tensor{r}{_A}\doteq \left(ar\right)\doteq \frac{1}{\sqrt{H^2+\kappa /a^2}},
\end{equation}
where $H$ is the Hubble parameter.
This corresponds to the location of the apparent horizon of the FLRW geometry \cite{Cai:2005ra, Faraoni:2011hf}.\\
The internal energy density is found by isometrically embedding $\mathbb{S}$ into an Euclidean 3-geometry and calculating its extrinsic curvature as $k_0=2/(ar)$.
Thus according to (\ref{eq:UwithFmin}) and (\ref{eq:internalateq}),
\begin{equation}
\mathpzc{U}\doteq \mathpzc{T}\,\mathpzc{S} \doteq \frac{1}{2}\frac{1}{\sqrt{H^2+\kappa /a^2}},
\end{equation}
with $\mathpzc{T}\doteq \lambda/(8\pi ar) \doteq \lambda\sqrt{H^2+\kappa /a^2}/(8 \pi)$,
and $\mathpzc{S}\doteq Area(\mathbb{S})/\lambda \doteq 4 \pi /\left[\lambda\left(H^2+\kappa /a^2\right)\right]$.

For $\lambda=4$, this result matches the one of \cite{Cai:2005ra,Jiang:2009zzf,Jiang:1900zza,Zhu:2010zzf}, where the temperature attributed to the apparent horizon is found to be $T_A=1/\left(2\pi r_A \right)$.

This temperature, assigned to the whole system, can be shown to be a worldsheet--constant by the variation
\begin{equation}
\delta \mathpzc{T}=\frac{1}{2}\sqrt{\tilde{\nabla}_a \left(\frac{\lambda}{8\pi ar}\right)\tilde{\nabla}^a \left(\frac{\lambda}{8\pi ar}\right)}\doteq 0.
\end{equation}
When this condition holds, the work term (\ref{eq:work}) and the Gibbs free energy (\ref{gibbsenergy}) are also found to be
\begin{equation}
\mathpzc{W}\doteq \frac{1}{16\pi }\sqrt{2\bigg| \frac{2\kappa +2a\ddot{a}+2\dot{a}^2}{a^2}\bigg|}\left(4\pi a^2r^2\right),
\qquad
\mathpzc{G}\doteq \frac{1}{16\pi }\sqrt{2\bigg|\frac{2\ddot{a}}{a}+\frac{2}{r^2a^2}\bigg|} \left(4\pi a^2r^2\right).
\end{equation}
By (\ref{eq:FLRWeqcond}),
\begin{equation}
\mathpzc{G}\doteq \mathpzc{W}\doteq \frac{1}{2} \sqrt{\bigg| \frac{\kappa +a\ddot{a}+\dot{a}^2}{a^2}\bigg|}\frac{1}{H^2+\kappa /a^2}
\end{equation}
so that when the Friedmann equations are inserted we obtain
\begin{equation}\label{eq:GibbsWorkFLRW}
\mathpzc{G}\doteq \mathpzc{W}\doteq \frac{1}{4}\sqrt{\bigg| \frac{1-3p/\rho}{\frac{4\pi \rho}{3}}\bigg|},
\end{equation}
where $p$ is the pressure, $\rho$ is the energy density of the perfect fluid and their ratio is $\omega=p/ \rho$.
Alternatively, we can compare the work required to create a quasilocal 2-surface that encloses a system filled with either vacuum energy ($\omega=-1$),
stiff matter ($\omega=1$), dust ($\omega=0$) or radiation ($\omega=1/3$).
For the same value of the perfect fluid energy density, at equilibrium, the results state
\begin{equation}
\mathpzc{W}_{Vacuum}>\mathpzc{W}_{Stiff}>\mathpzc{W}_{Dust}>\mathpzc{W}_{Radiation}=0,
\end{equation}
meaning that a system filled with stiff matter has a greater tendency to store the potential relative work than a system filled with dust or radiation.
Note that the surface tension is independent of the spatial size of the system in a FLRW spacetime, consistent with the fact that the FLRW geometry models a homogeneous universe.
To see the differences with an inhomogeneous universe one may consider the Lema\^{\i}tre-Tolman spacetime.
\subsection{Lema\^{\i}tre-Tolman (LT) geometry}
The LT metric can be written in the comoving coordinates as
\begin{equation}
ds^2=-dt^2+\left(\frac{R'^2(t,r)}{1+2K(r)}\right)dr^2+R^2(t,r)d \Omega^2,
\end{equation}
where $X'=\frac{\partial X}{\partial r}$ and $\dot{X}=\frac{\partial X}{\partial t}$ for an arbitrary function $X$.
One can choose a comoving observer dual dyad
\begin{eqnarray}
\tensor{E}{^\mu _{\hat{0}}}=\left(1,0,0,0\right), \qquad \qquad \qquad \, \tensor{n}{^\mu _{\hat{2}}}=\left(0,0,\frac{1}{R},0\right),\\
\tensor{E}{^\mu _{\hat{1}}}=\left(0,\frac{\sqrt{1+2K}}{R'},0,0\right), \qquad \tensor{n}{^\mu _{\hat{3}}}=\left(0,0,0,\frac{1}{R \sin \theta}\right).
\end{eqnarray}

Then the Helmholtz free energy density for such an observer is given by
\begin{equation}
\mathpzc{f}=\sqrt{2\left(\frac{2+4K-2\dot{R}^2}{R^2}\right)}.
\end{equation}
Minimizing $\FF$ to obtain the condition for quasilocal thermodynamic equilibrium we find
\begin{equation}
\dot{R}^2\doteq 1+2K,
\end{equation}
which again gives the location of the generalized apparent horizon of the LT geometry \cite{9780511617676}.
In the absence of the cosmological constant, the evolution equation for the LT spacetime may be written as
\begin{equation}\label{eq:LTevolution}
\dot{R}^2=2K+\frac{2M}{R},
\end{equation}
where $M=M(r)$ is an arbitrary function which is said to play the role of the active gravitational mass within a constant radius shell in LT solutions.
Therefore, another way of defining the apparent horizon is $R(t,r) \doteq 2M(r)$.
Then after computing the internal energy density as $k_0=2/R$ by (\ref{eq:UwithFmin}) and (\ref{eq:internalateq}), the internal energy becomes
\begin{equation}
\mathpzc{U}\doteq \mathpzc{T}\,\mathpzc{S} \doteq M(r),
\end{equation}
with $\mathpzc{T}\doteq \lambda/(8\pi R) \doteq\lambda/(16\pi M)$,
and $\mathpzc{S}\doteq Area(\mathbb{S})/\lambda \doteq (16 \pi M^2)/{\lambda}$.
If we take $\lambda=4$, then the temperature assigned to the system takes the same value as the temperature attributed to the apparent horizon in \cite{Biswas:2011gw}.
The work term (\ref{eq:work}) and the Gibbs free energy (\ref{gibbsenergy}) are also found as:
\begin{equation}\label{eq:WorkLT}
\mathpzc{W}\doteq \frac{1}{16 \pi}\sqrt{2\bigg|-\frac{2K'}{RR'}+\frac{2\ddot{R}}{R}+\frac{2\dot{R}\dot{R}'}{RR'}\bigg|}\left(4\pi R^2\right),
\end{equation}

\begin{equation}\label{eq:GibbsLT}
\mathpzc{G}\doteq \frac{1}{16 \pi}\sqrt{2\bigg|\frac{2\ddot{R}}{R}-\frac{2\dot{R}^2}{R^2}+\frac{2(2K+1)}{R^2}-\frac{\left(2K'-2\dot{R}\dot{R}'\right)}{RR'}\bigg|}\left(4\pi R^2\right).
\end{equation}\\
Substituting (\ref{eq:LTevolution}) and $R(t,r) \doteq 2M(r)$ into the equations above gives
\begin{equation}\label{eq:GibbsWorkLT}
\mathpzc{G}\doteq \mathpzc{W}\doteq M(r)/\sqrt{2} .
\end{equation}
In contrast to the homogeneous cosmology, the surface tension in
(\ref{eq:WorkLT}) depends on the radial position of the quasilocal observers
who define the inhomogeneous system.

One can check the thermodynamic potentials of the LT system reduce to those
of the FLRW and Schwarzschild spacetimes in the appropriate limit.
In particular, for \cite{9780511617676}
\begin{equation}
R=a(t)r \qquad and \qquad M=\int 4\pi \rho R^2 R' dr
\end{equation}
the relative work term (\ref{eq:GibbsWorkLT}) agrees with the dust case of
(\ref{eq:GibbsWorkFLRW}) for the FLRW geometry as expected.
Likewise, if we take $R=r$ and use the spatial derivative of the evolution
equation (\ref{eq:LTevolution}) to eliminate $K'(r)$ then at equilibrium
(\ref{eq:WorkLT}) agrees with (\ref{eq:WorkSch}) for the Schwarzschild geometry.
However, the general relative work (\ref{eq:GibbsWorkLT})
of LT differs from the Schwarzschild case (\ref{eq:GibbsWorkSch}),
on account of the competing terms in (\ref{eq:WorkLT}) or (\ref{eq:GibbsLT}).
Recall that the Gibbs free energy is a measure of how much potential the
system possesses to do work. In the static limit the second and third terms
inside the square root in (\ref{eq:WorkLT}) vanish. In that case, the
system stores all of the gravitational energy due to the spatial term
$-2K'/(RR')$ as potential work without being compelled to expend
some of this energy as the system evolves in time.
\section{Local versus quasilocal equilibrium}\label{seclocalvsqlocal}

In Newtonian physics a system composed of particles which are in local thermodynamic equilibrium with each other is not always expected to be in global thermodynamic equilibrium.
In the case of global equilibrium, one can assign single values of thermodynamic variables to the whole system.
Likewise, in our case we have defined a quasilocal thermodynamic equilibrium condition so that one can assign a single temperature value to the whole system.

In general, there is no reason for a system in local (or global) equilibrium to be in hydrodynamic equilibrium as well.
One requires additional conditions for them to coincide \cite{Green:2013ica},
so we should be careful to distinguish these concepts.

A typical example of the {\em local} thermodynamics of matter fields
on a curved background is given by
Gao \cite{Gao:2011hh} who generalizes early work of Sorkin, Wald and Zhang
\cite{Sorkin:1981wd} to a generic perfect fluid.
He investigates the connection between local thermodynamic equilibrium and
hydrostatic equilibrium.
Gao considers a collection of monatomic ideal gas particles with $p=p(T)$,
$\rho=\rho(T)$ and $s=s(\rho,n)$ where $s$ is the locally defined entropy
density and $n$ is the number density of the particles.
By maximizing the local matter entropy, an equation for local hydrostatic
equilibrium is obtained.
This is not a unique way of defining the local thermodynamic equilibrium but
this specific thermodynamic equilibrium condition also satisfies the local
hydrostatic equilibrium.
Also it is important to note that the fluid particles are not necessarily in
a local thermal equilibrium here.

Now let us consider the analogous problem of the conditions under which {\em quasilocal} thermodynamic equilibrium and {\em quasilocal} hydrodynamic equilibrium coincide for a general system containing both matter and gravitational energy contributions.
We will assume that locally defined condition given by Green, Schiffrin and
Wald \cite{Green:2013ica} for matter fields also holds in the quasilocal case.
This requires the quasilocally defined entropy to have its extremum value with respect to a total variation defined by (\ref{eq:firstlaw}).
In order to compute this one can consider a generic spherically symmetric
spacetime metric
\begin{equation}\label{eq:sphsymmmetric}
ds^2=-A^2(r,t)dt^2+B^2(r,t)dr^2+R^2(r,t)d\Omega^2,
\end{equation}
where $A(r,t)$ and $B(r,t)$ are arbitrary functions, $R(r,t)$ is the areal radius of $\mathbb{S}$, and our double dyad is
\begin{eqnarray}
\tensor{E}{^\mu _{\hat{0}}}=\left(\frac{1}{A(r,t)},0,0,0\right), \qquad \, \tensor{n}{^\mu _{\hat{2}}}=\left(0,0,\frac{1}{R(r,t)},0\right),\\
\tensor{E}{^\mu _{\hat{1}}}=\left(0,\frac{1}{B(r,t)},0,0\right), \qquad \tensor{n}{^\mu _{\hat{3}}}=\left(0,0,0,\frac{1}{R(r,t)\sin \theta}\right).
\end{eqnarray}
Then, quasilocal Helmholtz free energy takes its minimum value when $A^2R'^2 \doteq B^2\dot{R}^2$, and the total variation of the quasilocally defined entropy, (\ref{eq:entropy}), at equilibrium is
\begin{eqnarray}
\delta \mathpzc{S}&=&\frac{1}{2}\sqrt{\tilde{\nabla}_a\mathpzc{S}\tilde{\nabla}^a\mathpzc{S}}
\nonumber \cr
&=&\frac{1}{2}\sqrt{\tensor{E}{^\mu _{a}}\partial _{\mu}\left(\frac{4 \pi R^2}{\lambda}\right)\tensor{\eta}{^a^b}\tensor{E}{^\nu _{b}}\partial _{\nu}\left(\frac{4 \pi R^2}{\lambda}\right)}\doteq 0,
\end{eqnarray}
showing that the entropy is an extremum.
This allows us to conclude that the quasilocal thermodynamic equilibrium and quasilocal hydrodynamic equilibrium should coincide.
The interpretation of this result is crucial for the next section.
\section{Quasilocal virial relation} \label{secvirial}
Here we will sketch how the formalism above might be adapted to give a
quasilocal virial condition which differs in character from previous attempts
to define a virial theorem in general relativity \cite{1963Chandrasekar, 1973ApJ...182..335B, 1979ApJ...227..307V, 0264-9381-11-2-015}.
In previous studies only matter fields have been investigated, in which the
central object is the energy-momentum tensor defined at each spacetime point.
However, a full description of the virial theorem in general relativity
should also include gravitational energy, which
cannot be defined at a point due to the equivalence principle.

Ordinarily in classical mechanics the virial theorem is obtained by considering a system with motions confined to a finite region of space.
If the potential energy is a homogeneous function of the coordinates, then the virial theorem gives a relation between the time averaged values of the total kinetic and potential energies \cite{1969mech.book.....L}.
For such a system one can define a virial function $G$ by \cite{2002clme.book.....G} $G(t)=\sum _{i} \vec{p}_i.\vec{r}_i$,
where $\vec{r}_i$ are the coordinates and the $\vec{p}_i$ are the momenta of the particles in the system.
If $G(t)$ is a bounded function, then the mean value of its time derivative is zero, i.e.,
\begin{equation}
\bigg <\frac{d}{dt}\left(\vec{p}_i.\vec{r}_i\right)\bigg>=\bigg <\sum _{i}\left(\vec{p}_i.\vec{v}_i\right)\bigg>+\bigg <\sum _{i}\left(\vec{r}_i.\vec{\dot{p}}_i\right)\bigg>=0.
\end{equation}
Therefore, for a system under gravitational potential, the virial theorem reads
\begin{equation}
2\mean{K.E}=-\mean{P.E},\label{eq:newtonianvirial}
\end{equation}
where $\mean{K.E}$ and $\mean{P.E}$ are the time averaged values of the total kinetic and potential energies.

When it comes to including relativistic effects, however, this result changes.
For the ultrarelativistic limit $\left(v\rightarrow c\right)$, the virial theorem takes the form \cite{1978vtsa.book.....C}
\begin{equation}\label{eq:specrelvirial}
\mean{K.E}=-\mean{P.E}.
\end{equation}

In astrophysics, the virial theorem is used when the thermal and the gravitational forces acting on an isolated system balance each other so that the system neither expands nor contracts.
This state of the system is defined by its hydrostatic equilibrium \cite{1992isa..book.....B}.
In general the system is assumed to be composed of ideal gas particles which are in local thermal equilibrium with each other, which guarantees stability \cite{Green:2013ica}.
The value of the internal energy of the system, $E^{in}$, is then equal to the ensemble average of the kinetic energies of the particles creating the system \cite{2000itss.book.....P}, i.e.,
\begin{equation}
E^{in}=\overline{K.E},
\end{equation}
where the overbar denotes the ensemble average at a given time.
Note that the equipartition theorem is an application of the virial theorem.
If thermal equilibrium coincides with hydrostatic equilibrium, then the temporal average of the kinetic energy of the total system becomes equal to the ensemble average of the kinetic energies of the particles at a given time \cite{greiner2000thermodynamics}, i.e., $\mean{K.E}|_{\{t_1\rightarrow t_2\}}\equiv \overline{K.E}|_{\{t\}}$.
Consequently, for this case, one can rewrite (\ref{eq:newtonianvirial}) and (\ref{eq:specrelvirial}) as

\begin{align}\label{eq:newtonianvirialin}
2E^{in}=-\mean{P.E} \qquad(nonrelativistic)
\end{align}
and \cite{1978vtsa.book.....C}
\begin{align}\label{eq:specrelvirialin}
E^{in}=-\mean{P.E} \qquad(ultrarelativistic)
\end{align}

In our case, quasilocal thermodynamic equilibrium is set by the minimum of the Helmholtz free energy which holds for systems with worldsheet--constant temperature and volume.
This occurs when the mean extrinsic curvature of $\mathbb{S}$ is zero.
Thus if one perturbs $\mathbb{S}$ along $\mathbb{T}$, it \textit{neither expands nor contracts}.
Furthermore, the quasilocally defined entropy of the system then takes its
extremum value.
We can then expect the quasilocal thermodynamic and hydrodynamic equilibria
to coincide.

By analogy to the case of matter fields only, a
virial theorem might hold also in our quasilocal formalism.
In particular, we will suggest a natural bound,
\begin{equation}\label{eq:genbindingminfinal}
E_{matter+grav}\doteq E_{KLY}-E_{B}^{min}.
\end{equation}
from which a quasilocal virial relation follows. Here $E_{matter+grav}$
represents the combined energy of the matter and gravitational fields, while
$E_{B}^{min}$ is the minimum binding energy of the system. Furthermore,
the interpretation of $E_{KLY}$ as an invariant proper mass-energy for
generic spacetimes is
agreed upon by many authors \cite{Epp:2000zr,Liu:2003bx,Murchadha:2007em}.
We demonstrate that (\ref{eq:genbindingminfinal}) agrees with known results
in particular limits, but will not attempt a formal proof of this bound or
consequently of a virial theorem, which would require a detailed definition of
$E_{matter+grav}$ in terms of suitable ensemble averages.

We first recall that Bizon, Malec and \' O Murchadha \cite{Bizon:1990nk}
introduced a mass bound for a collection of spherical shells under collapse,
given by
\begin{equation}\label{eq:massineqfin}
M\leq M_p-E_B,
\end{equation}
where $M$ is the total energy of the shells, $M_p$ is the total proper mass
and $E_B$ is the binding energy. The equality holds when binding energy is
minimum, which turns out to be the Newtonian limit. Later, Yu and Caldwell
\cite{Yu:2008ij} included this argument in their calculation of the binding
energy of a Schwarzschild black hole and showed that
\begin{equation}
M=M_{p}-E_{B}^{min},\label{ycmin}
\end{equation}
and $M_p=E_{BY}$ in the Schwarzschild geometry for static observers at any $r$. 
Since $E_{BY}=E_{KLY}$ for static observers in the Schwarzschild geometry for
observers who are instantaneously at rest, (\ref{ycmin}) is seen to
coincide with (\ref{eq:genbindingminfinal}) in the Schwarzschild geometry.

Now let us specialize to observers at quasilocal hydrostatic equilibrium in
the Schwarzschild geometry, at $r=2M$, where the Kijowski-Liu-Yau energy gives
\begin{align}\label{eq:EKLYHydro}
E_{KLY} = \frac{-1}{8\pi}\int_{\mathbb{S}}{d\mathbb{S}\left(\frac{2\sqrt{1-\frac{2M}{r}}}{r}-\frac{2}{r}\right)} \doteq 2M.
\end{align}
Since the internal energy, $E_K^{in}$ (\ref{eq:E_Kin}), at equilibrium
corresponds to the usable matter + gravitational energy, the l.h.s. of
(\ref{eq:genbindingminfinal}) should read
\begin{eqnarray}
E_{matter+grav}&=&E_{K}^{in}\nonumber \cr
&&=\frac{-1}{16\pi}\int_{\mathbb{S}}{ d\mathbb{S}\left(\frac{\frac{4\left(1-\frac{2M}{r}\right)}{r^2}-\frac{4}{r^2}}{\frac{2}{r}}\right)}\doteq M.
\end{eqnarray}
Hence $E_B^{min}\doteq M$. The minimum binding energy of such a system can also be found by calculating the work done in bringing the spherical mass shells from infinity to $r=2M$ by observers who are instantaneously at rest in the Newtonian limit \cite{Yu:2008ij}.
Thus at hydrostatic equilibrium,
\begin{equation}\label{eq:virial}
E_{matter+grav}\doteq E_{B}^{min},
\end{equation}
which is analogous to the virial relation (\ref{eq:specrelvirialin}) in the
ultrarelativistic limit, since binding energy is negative of the potential energy, and only the so-called reference term survives in $E_{K}^{in}$ at quasilocal thermodynamic equilibrium.

Now we will generalize this result by {\em assuming} that
(\ref{eq:genbindingminfinal}) holds for {\em generic} spherically symmetric
spacetimes at quasilocal hydrodynamic equilibrium.
At quasilocal hydrodynamic equilibrium
$ \sqrt{2\left(\tensor{J}{_a^i^j}\tensor{J}{^a_j_i}\right)}=\sqrt{k^2-l^2}\doteq 0$, by (\ref{eq:E_Kin}) and (\ref{eq:E_K}),
\begin{equation}\label{Ekmingen}
E_{matter+grav}\doteq E_{K}^{in}\doteq \frac{1}{16\pi}\oint_{\mathbb{S}}{k_0 \, d\mathbb{S}}.
\end{equation}
\begin{equation}
E_{KLY}\doteq 2\left(\frac{1}{16\pi}\oint_{\mathbb{S}}{k_0 \, d\mathbb{S}}\right),
\end{equation}
Hence from \ref{eq:genbindingminfinal} we find
\begin{equation}\label{Ebmingen}
E_{KLY}-E_{matter+grav}\doteq E_{B}^{min}\doteq \frac{1}{16\pi}\oint_{\mathbb{S}}{k_0 \, d\mathbb{S}},
\end{equation}
where $k_0=2/R(r,t)$ is the extrinsic curvature of $\mathbb{S}$ when embedded in Euclidean 3-space and $R(r,t)$ is the areal radius of $\mathbb{S}$.
Therefore, by (\ref{Ekmingen}) and (\ref{Ebmingen}) one obtains
(\ref{eq:virial}) as a virial relation for any spherically symmetric distribution at quasilocal hydrodynamic equilibrium whose matter and gravitational contributions to the total content cannot be decoupled.

A key inference is that the proper mass--energy, usable matter--energy and the
binding energy of a system make most sense when referred to measurements made
by the same set of quasilocal observers.
Therefore the idea of comparing certain energy definitions for systems with different sizes in a given spacetime can lead to paradoxes.
For example, Frauendiener and Szabados argued that \cite{Frauendiener:2011rm}
\textit{``...if the quasi-local mass} ($E_{KLY}$) \textit{should really tend to the ADM mass as a strictly
decreasing set function near spatial infinity, then the Schwarzschild example shows that
the quasi-local mass at the event horizon cannot be expected to be the irreducible mass.''}
This is true simply because a system with a spatial size coinciding with the event horizon has different binding energy requirements than the one whose spatial size tends to infinity.
The latter one has zero binding energy, because no work has to be done by the system defined by quasilocal observers located already at infinity.
The authors continue with the statement
\textit{
``...there would have to be a closed 2-surface between the horizon and
the spatial infinity on which the quasi-local mass would take its maximal value. However,
it does not seem why such a (geometrically, and hence, physically) distinguished 2-surface
should exist.''}
Here we note that such a closed spatial 2-surface does exist with a location matching the apparent horizon.
It encloses a system whose quasilocal thermodynamic equilibrium coincides with quasilocal hydrodynamic equilibrium.
This might also serve as a physical interpretation for a generalized apparent horizon for the case when its location matches the one of a marginally outer trapped surface.
The outgoing null rays of a system enclosed by such a trapped surface do not tend to leave the system because systems in quasilocal thermal equilibrium simply do not radiate.

\section{Discussion}

When the equilibrium thermodynamics \textit{of horizons} was first introduced
in the 1970s \cite{PhysRevD.7.2333, PhysRevD.9.3292, Bardeen:1973gs}, the
quasilocal energy definitions that we have today were unknown.
It is now known that the physically relevant boundary Hamiltonian of general
relativity lies on a closed 2-dimensional spacelike surface, $\mathbb{S}$, of a
spacetime domain \cite{Brown:1992br,1997GReGr..29..307K}, which we call the
{\em screen}. In this paper we have focused on a spherically symmetric
\textit{system} enclosed by a screen, $\mathbb{S}$, as the central object
of gravitational thermodynamics rather than horizons.

Isolated systems are natural objects in classical thermodynamics. In
general relativity, however, no system can be totally decoupled from the rest
of the universe due to the nonlinear nature of the gravitational interaction.
The systems we consider in this paper have arbitrary size and are generally in
nonequilibrium with their surroundings. Only after quasilocal thermodynamic
equilibrium conditions are introduced does it follow that the screen is located
at the apparent horizon of \cite{Hayward:1993wb}, where the standard equilibrium
thermodynamic laws apply.

We believe that this approach may ultimately prove useful in general relativity,
since the issues associated with quasilocal gravitational energy on one hand,
or with gravitational entropy on the other, are generally studied in isolation.
In fact, the problem of gravitational entropy is so complex that often
researchers simply seek definitions in terms of geometric quantities which
are nondecreasing with time \cite{Wainwright:1983pg,Gron:2000dp, Clifton:2013dha}, giving rise to a ``second law'', without
directly investigating whether the entropies so defined obey any of the other
properties one might reasonably demand of a genuine
thermodynamic potential. The fact that the second law of classical thermodynamics can be viewed as a consequence of
entropy not being rigorously defined in nonequilibrium \cite{1978Sci...201..777P} is usually overlooked.

On account of the equivalence principle, statistical macroscopic properties
of the gravitational field are necessarily nonlocal. To interpret quasilocal
gravitational energy in terms of thermodynamic laws it is necessary to have
a measure of the ``work done'' by the tidal forces on the screen associated
with the quasilocal observers. For this reason, we have adapted the generalized
Raychaudhuri equation of an arbitrary dimensional worldsheet embedded in an
arbitrary dimensional spacetime \cite{Capovilla:1994yk}, to the special case
of a 2-dimensional timelike surface, $\mathbb{T}$, (orthogonal to $\mathbb{S}$
at every point), which we embed directly into 4-dimensional spacetime.

The mean extrinsic curvature of $\mathbb{S}$, that appears in the quasilocal
energy definitions \cite{1997GReGr..29..307K,Epp:2000zr,Liu:2003bx}, gives
the expansion of $\mathbb{S}$ when it is perturbed along $\mathbb{T}$. Degrees
of freedom living on $\mathbb{T}$ are , therefore, understood to
be those which describe the changes of thermodynamic potentials, while the
degrees of freedom living on the screen, $\mathbb{S}$, are required to
consistently define a system. Hence, in order to write the first law, the
total variations of the thermodynamic variables are taken along the 2-surface
$\mathbb{T}$ rather than variations along the integral curve of a single
vector.

It is known that quasilocal energy definitions which involve the mean
extrinsic curvature of $\mathbb{S}$ are invariant under radial boosts. In our
formalism this boost invariance holds for all thermodynamic potentials that
appear in the generalized Raychaudhuri equation (\ref{eq:ContCurlyCovJ}), once
the definitions (\ref{eq:fsquare}), (\ref{eq:wdensity}) and (\ref{eq:gibbsdensity}) are made. This is possible on account
of spherical symmetry.

Spacetime geometry is the source of gravitational energy while also
defining the relationships between the clocks and rulers that must be used
to investigate the system.
One might wonder how quasilocal thermodynamic relations could be treated
in spacetimes more general than spherically symmetric ones.
In the absence of
symmetries there are no unique definitions of energy in general
relativity. Conservation laws, by definition, require that some property does
not change. Therefore only with some symmetry condition can one define a
system consistently by keeping certain degrees of freedom constant.

Since gravitational energy and related macroscopic variables
are not local, whether any particular definition has utility depends
on the consistent definition of 2--surfaces, $\mathbb{S}$, which are best
adapted to the spacetime in question. Our spherically symmetric formalism
might be easily extended to situations which are approximately spherically
symmetric, in a perturbative scheme. Furthermore,
we believe that similar
reasoning to ours could also apply to spacetimes with other
symmetries, such as axial symmetry. In that case one should be able to introduce
additional quantities, which account for rotational energy, for example.
In such a case the first and third terms on the r.h.s.\ of the generalized
Raychaudhuri equation (\ref{eq:ContCurlyCovJ}) are nonzero, making a
thermodynamic interpretation considerably more complicated, and is left to
future investigations.

As an application of our formalism, we have sketched a natural bound involving
the quasilocal gravitational energy plus matter fields, which might suggest
a virial relation. To rigorously prove a virial theorem requires that we
have a proper understanding of the degenerate states of matter and gravitational
fields contained within the screen, $\mathbb{S}$, which are consistent with
the same worldsheet--constants on $\mathbb{S}$. Such an understanding of
course requires going far beyond the present paper, as it effectively
means probing fundamental questions related to the holographic interpretation which are important to both statistical and
quantum gravity.

Other questions for future work relate to the question of nonequilibrium
quasilocal gravitational thermodynamics. The Helmholz free energy, (\ref{eq:Helmholtz}),
is defined for all states of the system, whereas the other thermodynamic
potentials have only been defined at quasilocal thermodynamic equilibrium.
This is simply because equilibrium was defined by minimization of $\FF$,
which may not be the only way to define a useful quasilocal thermodynamic
equilibrium condition. Other types of equilibrium conditions could be applied
to the generalized Raychaudhuri equation.

For the case of thermodynamic nonequilibrium, the Raychaudhuri equation of the
worldsheet (\ref{eq:ContCurlyCovJ}) should still quantify the energy fluxes
into and out of the system. However, in that case, the existence of
thermodynamic variables is not guaranteed and their consistent definition
becomes murky even in classical thermodynamics. Losing the linearity condition
among the thermodynamic variables makes their interpretations much more
difficult and one might need additional geometrical structures to investigate
processes involving their generation and dissipation.

Recently, Freidel \cite{Freidel:2013jfa} presented an approach to study
nonequilibrium thermodynamics by using geometrical objects in a 2+2 formalism.
Our language is similar to his in terms of the quasilocal nature of the
thermodynamic potential densities introduced on a screen $\mathbb{S}$. However,
our formalism differs fundamentally in character by the existence of a
worldsheet $\mathbb{T}$ on which both of the degrees of freedom are treated
equally in terms of their roles in evolving the potentials. Whether or not
investigation of this difference provides a passage from quasilocal
equilibrium to nonequilibrium thermodynamics is a point of interest.

In conclusion, we have introduced a new formalism to link two areas which
are often treated separately -- quasilocal energy and horizon thermodynamics --
by adapting a geometrical approach of Capovilla and Guven
\cite{Capovilla:1994yk} to define a quasilocal equilibrium condition.
We have investigated specific examples and potential
applications in the case of spherical symmetry. This already sheds fresh light
on some paradoxes relating to quasilocal energy. Although our paper is by
nature exploratory, we believe that the steps taken here can be further built
upon to potentially tackle important questions that generally are not
currently considered, but should be, in gravitational physics.
\section*{References}
\bibliographystyle{iopart-num}
\bibliography{references}
\end{document}